\documentstyle[12pt,epsf,epsfig]{article}
\newcommand{\bp}{{\bf p}}
\newcommand{\bq}{{\bf q}}
\newcommand{\bk}{{\bf k}}
\newcommand{\bx}{{\bf x}}
\newcommand{\by}{{\bf y}}

\newcommand{\btab}{\begin{tabbing}}
\newcommand{\etab}{\end{tabbing}}
\newcommand{\eqntimes}{\mbox{} \times}

\newcommand{\beqn}{\begin{equation}}
\newcommand{\eeqn}{\end{equation}}
\newcommand{\barr}[1]{\begin{array}{#1}}
\newcommand{\earr}{\end{array}}
\newcommand{\beqna}{\begin{eqnarray}}
\newcommand{\eeqna}{\end{eqnarray}}
\newcommand{\btablec}{\begin{table} \begin{center}}
\newcommand{\etablec}{\end{center} \end{table}}

\newcommand{\gapproxeq}{\lower.7ex\hbox{$\;\stackrel{\textstyle>}
{\sim}\;$}}
\newcommand{\plabel}[1]{\label{#1}}
\newcommand{\pbibitem}[1]{\bibitem{#1}}
\def\question#1{}
\input epsf

\begin{document}
\title{
\begin{flushright} 
\small{hep-ph/0303170} \\ 
\small{LA-UR-03-1658} 
\end{flushright} 
\vspace{0.5cm}  
\Large\bf Selection rules for $J^{PC}$ Exotic Hybrid Meson Decay in 
Large-$N_c$}
\vskip 0.2 in
\author{Philip R. Page\thanks{\small \em E-mail:
prp@lanl.gov}\\{\small \em  Theoretical Division, MS B283, Los Alamos
National Laboratory, Los Alamos, NM 87545, USA}}
\date{}
\maketitle
\begin{abstract}{
The coupling of a neutral hybrid
$\{1,3,5\ldots\}^{-+}$ exotic particle (or current) to two neutral 
(hybrid) meson particles
with the same $J^{PC}$ and $J=0$ is proved to be sub-leading to the usual 
large-$N_c$ QCD counting. The coupling 
of the same exotic particle to certain two - (hybrid) meson currents
with the same $J^{PC}$ and $J=0$ is also sub-leading.
The decay of a $\{1,3,5\ldots\}^{-+}$ hybrid particle
to $\eta\pi^0,\; \eta'\pi^0,
\eta'\eta,\; \eta(1295)\pi^0,$ $\pi(1300)^0\pi^0,$ $\eta(1440)\pi^0,$ 
$a_0(980)^0\sigma$ or $f_0(980)\sigma$ is sub-leading, assuming that
these final state particles are (hybrid) mesons in the limit of
large $N_c$.}
\end{abstract}
\bigskip


PACS number(s): 11.15.Pg \hspace{.16cm}11.15.Tk \hspace{.16cm} 
11.30.Na \hspace{.16cm}11.55.-m
\hspace{.16cm} 12.38.Aw \hspace{.16cm} 12.39.Mk \hspace{.16cm} 13.25.Jx

\vspace{.3cm}

\section{Introduction \label{sec1}}

States of Quantum Chromodynamics (QCD) can definitively be said not
to be conventional mesons
when these states have exotic $J^{PC}$,
which cannot be constructed
for conventional mesons in the quark model, or equivalently, cannot
be built from local currents with only a quark and an antiquark field.
Here $J$ denotes the internal angular momentum,
$P$ (parity) the reflection through the origin and 
$C$ (charge conjugation) particle-antiparticle exchange.
These are conserved quantum numbers of QCD.

With the experimental discovery of isovector $J^{PC}$ exotics, the
question of their interpretation has come into focus. QCD with a large
number of colours $N_c$ offers a systematic expansion in $1/N_c$
with considerable phenomenological success~\cite{lebed,coleman}, 
which can address this question. This is because a glueball (built
from only gluons) and a (hybrid) 
meson (quark-antiquark with additional gluons)
do not mix in large-$N_c$~\cite{lebed}. Furthermore, four-quark states 
(two quark-antiquark pairs) are absent~\cite{coleman}.
In large-$N_c$ the isovector $J^{PC}$ exotics {\it must} therefore be
hybrid mesons, as glueballs are isoscalar. 
Here it is proved for the first time that certain decays
of hybrid mesons that are allowed by the conserved quantum
numbers of QCD are sub-leading to their usual large-$N_c$ counting,
providing a consistency check for the hybrid nature of the state.

Selection rules for $J^{PC}$ exotic hybrid decay {\it amplitudes},
e.g. the amplitude for
$J^{PC}=1^{-+}$ hybrid particle $\to \eta\pi,\, \eta'\pi$,
were noticed in non-field theoretic analyses~\cite{sel}. 
In QCD it was found that
these selection rules are really properties of 
certain three-point {\it Green's 
functions}~\cite{pene,fssr}. The first attempt
to obtain hadronic properties from the
Green's functions~\cite{pene} contained some errors~\cite{fssr}.
These properties were subsequently extracted in finite-$N_c$ QCD,
e.g. the physical $N_c=3$~\cite{fssr}.
The properties were of limited physical relevance since they
pertained to the coupling of currents to particles, e.g.
a $1^{-+}$ hybrid current to $\eta\pi$. 
Also, for technical reasons, the scope of the deductions was limited.
As will be seen below, 
these reasons disappear in large-$N_c$, because three and
more particles do not contribute.
The large-$N_c$ treatment of the results of Ref.~\cite{fssr} is the
subject of this Paper. The results of this Paper can be comprehended
by only reading Section~\ref{sec4}, which also explicates the
experimental consequences of this Paper. Section~\ref{sec2}
proves two results for the coupling of particles to currents by
first proving two preliminary results.
Section~\ref{sec3} uses the results of the previous section
to prove a result for the coupling of particles
to particles.

\section{Coupling of currents to particles~\label{sec2}}

The decay of a $J^{PC}=\{1,3,5\ldots\}^{-+}$ particle to two
identical $J=0$ particles vanishes by Bose symmetry, because the
final state particles are in an odd partial wave. The analogous
statement for a Green's function built from a 
$J^{PC}=\{1,3,5\ldots\}^{-+}$ current and two $J=0$ currents
is that the ``identical current'' part, or symmetric part, of
the Green's function vanishes~\cite{fssr}. The OZI rule allowed
contributions to the Green's function only has a symmetric 
part~\cite{fssr}, so that they do not contribute to the
Green's function. The expression that does not contain an
OZI rule allowed contribution is

\beqna\plabel{main}\lefteqn{\hspace{-14.2cm}\int^{\infty}_{-\infty} dt\; 
e^{iEt}\; \hat{O}_{\bp}\;    
\int\; d^3x\; d^3y\; e^{\, i\,(\bp\cdot\bx-\bp\cdot\by)}\;
\langle  0 |\: B(\bx,t)\:C(\by,t) \:A_{\mu}(0)\: |0\rangle \nonumber } \\
=  \sum_n\; (2\pi)^4\; 
\delta^3(\bp_n)\; \delta(E_n-E)\;
\hat{O}_{\bp}\;\langle  0 |\: (\int d^3x\; 
e^{i\bp\cdot\bx}\; {B}(\bx,0){)} \; {C}(0)\:
|n\rangle \; \langle n| {A_{\mu}}(0)|0\rangle \; .
\eeqna
This is proved in Eqs. 2, 3 and 14 of Ref.~\cite{fssr} with no 
approximations. 
The left-hand side (L.H.S.) of the equation contains the time integral
and spatial Fourier transform of a three-point Green's function which
describes the ``decay'' of $A$ into $B$ and $C$. 
The expression is in Minkowski (physical) 
space with $E$ and $\bp$ real numbers.
The L.H.S. is 
expanded on the right-hand side (R.H.S.) by inserting an infinite set
of asymptotic stable states $n$ with energy $E_n$ and momentum $\bp_n$
in order to extract physical predictions. 
The delta functions indicate that the asymptotic states are at rest and
have energy $E$.
The gauge-invariant local currents $B$, $C$ and $A_{\mu}$ have the 
flavour structure of a neutral 
(hybrid) meson (linear 
combinations of $\bar{u}u,\bar{d}d,\ldots$ quark fields), 
and can contain gluon fields~\cite{restrict}. 
The currents $B$ and $C$ both have 
the same colour-Dirac-derivative-gluon structure for a given 
flavour, a finite number of
derivatives (when expanded as a power series) and $J=0$.
Also, the currents $B(0)$ and
$C(0)$ have equal $P$ and $C$.
The current $A_{\mu}(0)$ is assumed to have $P=-$ and odd $J$ (with
Lorentz indices denoted by $\mu$). 
Conservation of charge conjugation then implies that this current
is $J^{PC}=\{1,3,5\ldots\}^{-+}$ exotic, so that it should contain
at least one gluon field: a hybrid meson current.

Eq.~\ref{main} would be of limited interest were it not for the fact that
the action of the operator $\hat{O}_{\bp}$ 
(containing a finite number of derivatives in powers of $\bp$)
allowed the demonstration that the L.H.S. contains only OZI rule forbidden 
contributions, and hence is ${\cal O}(1)$ to leading order
in the large-$N_c$ power counting, as opposed to the usual 
${\cal O}(N_c)$. The remainder of this Paper exploits this behaviour
of the L.H.S. and deduces the consequences for the R.H.S. The strategy
is to keep only 
the leading contributions to the R.H.S. in large-$N_c$,
and then to equate to the L.H.S.
It is shown in the remainder of this section that the leading 
contribution to the R.H.S. comes from either 
$J^{PC}=\{1,3,5\ldots\}^{-+}$ one-hybrid-meson states, or from
two- (hybrid) meson states with $J=0$ and the same $J^{PC}$ (Eq.~\ref{fk}).
The R.H.S. contains a product of two matrix elements for each of the
leading contributions. One of the matrix elements will be shown to have
the usual large-$N_c$ counting. The other matrix element will 
be shown to have one order in $N_c$ lower counting than usual
(Eqs.~\ref{result2}-\ref{result1}), in order that the L.H.S. has an
order lower counting than usual, as required.

On the R.H.S. the usual large-$N_c$ counting for
$\langle  0 | B(\bx,0) \; C(0) |n\rangle$ is order
$\sqrt{N_c}$, $N_c$, $\sqrt{N_c}$ or $1$ for $n$ respectively a
one -, two -, three - or 
four - (hybrid) meson asymptotic state~\cite{lebed,cohen}.
The counting for 
$\langle n| {A_{\mu}}(0)|0\rangle$
is respectively $\sqrt{N_c}$, $1$, $1/\sqrt{N_c}$ or 
$1/N_c$~\cite{lebed,cohen}.
The product of the countings of the two matrix elements is
$N_c$, $N_c$, 1 and $1/N_c$ respectively.
If the asymptotic states contained glueballs the counting of the product
will be lower than $N_c$.
Hence only one- and two-particle (hybrid) meson
states contribute in large-$N_c$, and they contribute at ${\cal O}(N_c)$,
as they should to equal the usual counting of the L.H.S.
Also, the one-particle states that contribute to 
$\langle n| {A_{\mu}}(0)|0\rangle$ at ${\cal O}(\sqrt{N_c})$ are 
{\it only} neutral hybrid mesons with the same $J^{PC}$ as
the current $A_{\mu}(0)$~\cite{cohen}. These states cannot be mesons
because they are $J^{PC}$ exotic. The two-particle states
that contribute to 
$\langle  0 | B(\bx,0) \; C(0) |n\rangle$ at ${\cal O}(N_c)$
are {\it only} two neutral (hybrid) mesons~\cite{coleman}
with the same $J^{PC}$ as
the currents $B(0)$ and $C(0)$.
It follows that only one-hybrid-meson and two - (hybrid) meson
states contribute on the R.H.S. to leading order in large-$N_c$.
Using this the R.H.S. of Eq.~\ref{main} can be simplified to read
(Appendix~\ref{appa1})

\[
2\pi\sum_\sigma\delta(m_\sigma-E)\;\hat{O}_{\bp}\;\langle  0 |\: 
(\int d^3x\; 
e^{i\bp\cdot\bx}\; {B}(\bx,0){)} \; {C}(0)\:
|\sigma\,{\bf 0}\rangle \; \langle \sigma\,{\bf 0}| {A_{\mu}}(0)|0\rangle 
+\frac{1}{(2\pi)^2}\sum_{\sigma_1 \sigma_2} 
(1-\frac{\delta_{\sigma_1\sigma_2}}{2}) \nonumber
\]

\vspace{-1.0cm}
\beqn
\left.
\times K(E) \int d\Omega_{{\bf k}_1} 
\;\hat{O}_{\bp}\;\langle  0 |\: (\int d^3x\; 
e^{i\bp\cdot\bx}\; {B}(\bx,0){)} \; {C}(0)\:
|\sigma_1k_1 \sigma_2k_2\rangle \; 
\langle \sigma_1k_1 \sigma_2k_2| {A_{\mu}}(0)|0\rangle 
\;\right|_{k_1+k_2=({\bf 0},E)}  , \plabel{fk}
\eeqn
where the first sum is over the one-hybrid-meson states $\sigma$
(implicitly including the different polarizations); and
the second over the two - (hybrid) meson
states $\sigma_1$ and $\sigma_2$, in such a way that
a particular two-particle state 
is summed over only in the permutation
$\sigma_1\sigma_2$ (and not $\sigma_2\sigma_1$) in order to avoid 
double counting.
Throughout the text 
the convention is that non-boldfaced variables starting
with $k$ (and $p,q,x,y,z$)  indicate four-vectors.

The kinematical variable dependence of the one-particle terms in
Eq.~\ref{fk} is only on $E$ and $\bp$. If the hybrid particles have a
discrete spectrum, there would only be contributions for discrete
values of $E$ when $E=m_\sigma$. Because of the constraint
$k_1+k_2=({\bf 0},E)$, the two-particle terms also only depend on the
kinematical variables $E$ and $\bp$. Because $K(E)$ only has support
above the two-particle threshold, the two-particle terms all vanish below
the lowest threshold, and contains a (continuous in $E$) contribution
from each two-particle state above its threshold.

In each of the terms in Eq.~\ref{fk}, one of the matrix elements has
the usual large-$N_c$ counting.  For the one-particle terms this is
$\langle \sigma\,{\bf 0}| {A_{\mu}}(0)|0\rangle$. The reason is that,
barring accidental cancellations, this matrix element {\it has} to have
exactly its usual large-$N_c$ counting (${\cal O}(\sqrt{N_c})$), in
order not to violate the counting for two-point Green's functions (the
{\it first preliminary} proved in Appendix~\ref{appa2}). For the
two-particle terms in Eq.~\ref{fk}, $\langle 0 |\, B(\bx,0) \: C(0)\,
|\sigma_1k_1 \sigma_2k_2\rangle$ has the usual large-$N_c$ counting
(${\cal O}(N_c)$). The reason is that, barring accidental
cancellations, it {\it has} to be exactly ${\cal O}(N_c)$, in order
not to violate the counting for four-point Green's functions (the {\it
second preliminary} proved in Appendix~\ref{appa3}).

Eqs.~\ref{main}-\ref{fk} can schematically be written as L.H.S. $=$
R.H.S. $=\sum_m a_m b_m + \sum_m c_m d_m$, where the sums are over
$\sigma$ for the one-particle terms $\sum_m a_m b_m$ and over
$\sigma_1\sigma_2\Omega_{{\bf k}_1}$ for the two-particle terms $\sum_m
c_m d_m$. (As shown in Appendix~\ref{appb}, the integral over
$\Omega_{{\bf k}_1}$ can be written as a finite sum due to a partial
wave expansion).  The sum over $\sigma$ is finite, since the
one-particle states must have mass equal to $E$. The
sum over $\sigma_1\sigma_2$ is finite because there is a finite number
of two-particle thresholds below $E$. If the L.H.S. is subleading in
the large-$N_c$ counting, the R.H.S. is, but this does $\it not$ imply
that $a_m b_m$ is subleading, nor that $c_n d_n$ is, because it is
possible that both $a_m b_m$ and $c_m d_m$ have the usual large-$N_c$
counting, and there is a cancellation between the terms yielding a
subleading R.H.S. The arguments in Appendix~\ref{appb} have the
consequence that $a_m b_m$ and $c_m d_m$ are {\it each} subleading,
i.e. ${\cal O}(1)$. This follows by evaluating the equation
L.H.S. $=\sum_m a_m b_m + \sum_m c_m d_m$ multiple times for different
currents, so that a matrix equation is obtained which is then
inverted. This can only be done if the number of terms on the
R.H.S. is finite, as it is here.  If three- or more-particle terms
contributed on the R.H.S., as would generally be the case for the
physical $N_c=3$, it is not clear that the number of terms on the
R.H.S. is finite, and that the inversion can be done. For this reason,
the fact that three or more particles are subleading in the
large-$N_c$ expansion, is important to make progress in the
derivation. The remainder of this section is devoted to two results,
which although exhaustively derived in Appendix~\ref{appb}, are mostly
evident at this point.

The first preliminary ($\langle \sigma\,{\bf 0}|
{A_{\mu}}(0)|0\rangle$ {\it has} to be ${\cal O}(\sqrt{N_c})$) means
that $b_m$ is ${\cal O}(\sqrt{N_c})$.  Together with $a_m b_m$ is
${\cal O}(1)$, this implies that $a_m$ is ${\cal O}(1/\sqrt{N_c})$,
which yields most of the {\it first result} of the Paper that the
coupling of currents to a particle

\beqn\plabel{result2}
\langle  0 |\; B(\bx,t)\;C(\by,t) \; | \sigma\,{\bf 0} \rangle  =
{\cal O}(\frac{1}{\sqrt{N_c}})\; ,
\eeqn
where its usual counting is ${\cal O}(\sqrt{N_c})$.
This holds for a neutral on-shell hybrid 
meson particle $\sigma$ at rest with 
$J^{PC}=\{1,3,5\ldots\}^{-+}$. Also,
$B$ and $C$ are neutral gauge-invariant local (hybrid) meson 
currents at space-time positions $x$ and $y$ at equal time  
with flavour structure a linear combination of 
$\bar{u}u,\; \bar{d}d, \ldots$,
the same colour-Dirac-derivative-gluon structure for a given 
flavour, a finite number of
derivatives and $J=0$. The currents $B(0)$ and
$C(0)$ should have equal $P$ and $C$.

The second preliminary ($\langle 0 |\, B(\bx,0) \: C(0)\, |\sigma_1k_1
\sigma_2k_2\rangle$ {\it has} to be ${\cal O}(N_c)$) yields most of
the fact that $c_m$ is ${\cal O}(N_c)$. Since $c_m d_m$ is ${\cal
O}(1)$, this implies that $d_m$ is ${\cal O}(1/{N_c})$, which yields
most of the {\it second result} that the coupling of particles to a
current

\beqn\plabel{result1}
\langle \sigma_1k_1 \sigma_2k_2| {A_{\mu}}(z)|0\rangle = {\cal
O}(\frac{1}{N_c})\; , \eeqn 
where its usual counting is ${\cal O}(1)$. This holds for neutral
on-shell (hybrid) meson particles $\sigma_1$ and $\sigma_2$ with
identical $J^{PC}$ and $J=0$, and with arbitrary four-momenta $k_1$
and $k_2$. Also, $A_{\mu}(z)$ is a neutral gauge-invariant local
hybrid meson $J^{PC}=\{1,3,5\ldots\}^{-+}$ current with Lorentz
indices $\mu$ at space-time position $z$ with flavour structure a
linear combination of $\bar{u}u,\; \bar{d}d, \ldots$.

\section{Coupling of particles to particles~\label{sec3}}

In this section the dependence on the current $A_\mu(z)$ is removed
from the matrix element in Eq.~\ref{result1} to obtain a result
(Eq.~\ref{result3}) that does not depend on the current, but on the
physically relevant T-matrix. Because the matrix element in 
Eq.~\ref{result1} has one order in $N_c$ lower counting than the
usual, the T-matrix element (Eq.~\ref{result3}) will have an order
lower counting than usual. It is shown in Appendix~\ref{appa4} that

\beqna
\lefteqn{\hspace{-13.7cm}\langle \sigma_1k_1 \sigma_2k_2\,\mbox{out}\,|
{A_{\mu}}(z)|0\rangle
-\langle \sigma_1k_1 \sigma_2k_2\,\mbox{in}\,|
{A_{\mu}}(z)|0\rangle  =
\sum_\sigma \left[ \frac{i 2\pi}{\sqrt{N_c}} 
\langle \sigma \; (k_1+k_2)\,\mbox{in} \, |{A_{\mu}}(z)|0\rangle
\right. \nonumber } \\
\eqntimes \left.\delta(\sqrt{m_1^2+\bk_1^2}+\sqrt{m_2^2+\bk_2^2}
-m_\sigma\,)\; \right] \left[ \sqrt{N_c}\,
\langle \sigma_1k_1 \sigma_2k_2 |\:  T\:  |
\sigma \; (k_1+k_2) \rangle\right] \; 
\plabel{en} 
\eeqna
when the restriction to the rest frame, $\bk_1+\bk_2={\bf 0}$, applies.
This equation states that when the difference of the coupling of
remote future ``out'' states and remote past ``in'' states to the current
is considered, valuable information about the physically relevant
T-matrix is obtained: the {\it third result} that the coupling of a
particle to particles

\beqn\plabel{result3}
\langle  \sigma_1k_1 \sigma_2k_2|\: T\:
|\sigma\, {\bf 0} \rangle = 
{\cal O}(\frac{1}{N_c^\frac{3}{2}}) \; ,
\eeqn
where its usual counting is ${\cal O}(1/\sqrt{N_c})$.
This holds for neutral
on-shell (hybrid) 
meson particles $\sigma_1$ and $\sigma_2$ with 
$J=0$, identical $J^{PC}$, and four-momenta $k_1$ and $k_2$
in the rest frame $\bk_1+\bk_2  = {\bf 0}$;
and for a neutral on-shell hybrid 
meson particle $\sigma$ at rest with 
$J^{PC}=\{1,3,5\ldots\}^{-+}$.

Even though the third result is proved in Appendix~\ref{appb} using
techniques analogous to those used to derive the first two
results, its plausibility can be verified by using Eq.~\ref{result3}
in conjunction with the first preliminary ($\langle \sigma\, {\bf
0}|{A_{\mu}}(z)|0\rangle$ {\it has} to be ${\cal O}(\sqrt{N_c})$), to
obtain that the R.H.S. of Eq.~\ref{en} is ${\cal O}(1/N_c)$,
consistent with the L.H.S. given by the second result
(Eq.~\ref{result1}) as ${\cal O}(1/N_c)$.

\section{Remarks~\label{sec4}}

The three results of the Paper are Eqs.~\ref{result2},
\ref{result1} and \ref{result3}, including the
discussion under each equation. 
These results are theorems of large-$N_c$ QCD field theory
with no approximations, and are valid 
within the generic large-$N_c$ 
framework~\cite{lebed,coleman,cohen}.

The first result (Eq.~\ref{result2}) implies that certain four-quark currents
are not good interpolators for hybrid meson particles. This may have
implications for Euclidean space lattice QCD, even though the result
was derived only in Minkowski space. A special case of the second result
(Eq.~\ref{result1}) was previously derived~\cite{fssr} for an
$\eta\pi^0$ asymptotic state for certain quark masses within a certain
kinematical range.

The third result (Eq.~\ref{result3}) is of direct experimental relevance.
For {\it example}, the decay amplitudes (couplings of a particle to particles)
of a $\{1,3,5\ldots\}^{-+}$ 
hybrid to $\eta\pi^0,\; \eta'\pi^0,
\eta'\eta,$ $\eta(1295)\pi^0,\; \pi(1300)^0\pi^0,$ $\eta(1440)\pi^0,$ 
$a_0(980)^0\sigma$ or $f_0(980)\sigma$ are ${\cal O}(1/N_c^\frac{3}{2})$, 
while the usual counting is ${\cal O}(1/\sqrt{N_c})$,
assuming that these final state particles are (hybrid) mesons in the limit of
large $N_c$. Hence the widths of these decays are $1/N_c^2$ suppressed
with respect to their usual counting. This is the same suppression that
large-$N_c$ predicts for decays forbidden by the OZI rule~\cite{lebed},
implying that the suppressions predicted here should be
similar phenomenologically.
The selection rule is most useful when 
OZI allowed decay is expected to be important in the absence of the selection
rule. In the example above, this is true for a hybrid
composed dominantly of $u\bar{u}$ and $d\bar{d}$. An the other hand,
it can be deduced from Eq.~\ref{result3} that the
coupling of a $1^{-+}$ hybrid to $\eta_c\eta$ is 
${\cal O}(1/N_c^\frac{3}{2})$, but this is less useful as the 
OZI allowed coupling (of the $c\bar{c}$ component of the 
hybrid to $\eta_c$ ($c\bar{c}$) and
the $c\bar{c}$ component of the $\eta$) is not expected to be important.
Interestingly, even in the
unlikely case where the $\eta'$ or $\sigma$ is a pure glueball 
in the limit of large $N_c$, the decay amplitude of $\{1,3,5\ldots\}^{-+}$ 
hybrids to $\eta'\pi^0, \eta'\eta,$ $a_0(980)^0\sigma$ or $f_0(980)\sigma$
would be ${\cal O}(1/N_c)$~\cite{lebed}, which is still subdominant to the
usual counting.
In case $\sigma$ is a meson-meson state in the
limit of large $N_c$, the predictions mentioned do not apply.
Beside the $0^{-+}$ and $0^{++}$ particles mentioned,
examples can also be given of $0^{+-}$ and $0^{--}$ exotic particles 
in the final states. 
The large-$N_c$ selection rules, in contrast to the selection
rules discussed in Section~\ref{sec1}, also apply when both
final state mesons do not have the same radial excitation, e.g
the $\eta(1295)\pi^0,\; \pi(1300)^0\pi^0$ and $\eta(1440)\pi^0$
final states.
Assuming isospin symmetry the results can also be extended to
charged states by use of the Wigner-Eckart theorem, as will now be done.

Consider the decay of a $1^{-+}$ isovector hybrid
with isospin symmetry. Decay to
$\eta\pi,\; \eta'\pi,$ $\eta(1295)\pi,$ $\eta(1440)\pi$ and
$a_0(980)\sigma$, which is ordinarily important, is suppressed.
The experimental 
$\pi_1(1600)$~\cite{pdg02} is a $1^{-+}$ exotic isovector resonance.
It has not been seen in $\eta\pi$.
A $1^{-+}$ enhancement at 1.6 GeV has prominently been seen in 
$\eta'\pi$~\cite{pdg02}, although the branching ratio is not dominant
if the enhancement is resonant
($B(\pi_1(1600)\to f_1\pi) / B(\pi_1(1600)\to \eta'\pi) =
3.80 \pm 0.78$~\cite{kuhn}). If the enhancement is dominantly 
non-resonant, as has been advocated~\cite{sz}, the branching ratio
is very small.
The decay $\pi_1(1600) \to \eta(1295)\pi$ is found to be small relative
to $f_1\pi$ in an analysis of the $\eta\pi^+\pi^-\pi^-$
final state~\cite{kuhn}, although an earlier report stated
that $\pi_1(1600)$ was seen in $f_1\pi$ and 
$\eta(1295)\pi$ at a similar magnitude in 
$K^+ \bar{K}^0\pi^-\pi^-$~\cite{bnl}. 
If $\pi_1(1600)$ is found to have a
large branching ratio to $\eta'\pi$, that would be inconsistent with
large-$N_c$ expectations which are otherwise consistent with 
its being a hybrid meson~\cite{qcdsr}. As discussed above, 
decay to $\eta'\pi$ is large-$N_c$ suppressed when $\eta'$ is
either a meson or a glueball in the limit of large $N_c$, although
the suppression is less when $\eta'$ is a glueball.
Hence a sizable $\eta'\pi$ branching ratio can arise through  
a large glueball component of an  
$\eta'$ meson~\cite{sel}, which violates the large-$N_c$ prediction
that meson-glueball mixing is suppressed.
The recently discovered $\pi_1(2000)$ has not been seen in 
$\eta\pi$, $\eta'\pi$ and $\eta(1295)\pi$~\cite{kuhn}, consistent with
its being a hybrid meson.



This research is supported by the Department of Energy under contract
W-7405-ENG-36.

\appendix
\section{Appendix: Diverse results}

\subsection{Derivation of Eq.~\ref{fk}\plabel{appa1}}

For the one-particle states,
$\sum_n = \sum_\sigma\int d^3 p_\sigma \,/\,(2\pi)^3$,
and the momentum ${\bf p}_\sigma =
\bf 0$ due to the momentum $\delta$-function,
so that $E_\sigma = \sqrt{m^2_\sigma+{\bf p}_\sigma}=m_\sigma$
because the particles in an asymptotic state are on-shell.
For the two-particle states
$\sum_n = \sum_{\sigma_1\sigma_2}(1-\delta_{\sigma_1\sigma_2}/2)\:
\int d^3 k_1 \,
/\,(2\pi)^3$ $\int d^3 k_2\, /\,(2\pi)^3$, where the factor
$(1-\delta_{\sigma_1\sigma_2}/2)$ is $1/2$ for the phase space of
identical particles. Substituting  
$\bp_n = \bk_1+\bk_2$ and 
$E_n =\sqrt{\bk_1^2+m_1^2}+\sqrt{\bk_2^2+m_2^2}$ 
in the phase space integration 

\beqn
\int \frac{d^3 k_1}{(2\pi)^3} \int \frac{d^3 k_2}{(2\pi)^3}
\; (2\pi)^4\; \delta^3(\bp_n)\; \delta(E_n-E)\; f(k_1,k_2)
= \left. \frac{K(E)}{(2\pi)^2}  \int d\Omega_{{\bf k}_1} \;
f(k_1,k_2) \;\right|_{k_1+k_2=({\bf 0},E)},
\eeqn
where
\beqna
\hspace{-0.5cm} 
K(E) & \equiv & \int^\infty_0 \bk_1^2\, 
d |\bk_1| \;\delta(\sqrt{\bk_1^2+m_1^2}+
\sqrt{\bk_1^2+m_2^2}-E) \nonumber \\
& = & 
\frac{1}{8E^4}\; (E^4-(m_1+m_2)^2(m_1-m_2)^2)
\;\sqrt{(E^2-(m_1+m_2)^2)(E^2-(m_1-m_2)^2)}\plabel{ke}
\eeqna
if $E \geq m_1+m_2$; and $K(E)$ vanishes if $E < m_1+m_2$. 

\subsection{First preliminary \plabel{appa2}}

The following two-point function is ${\cal O}(N_c)$ in the 
Feynman-diagrammatic large-$N_c$ counting~\cite{cohen}

\beqn
\langle 0|A_\mu(x_1)A_\nu(x_2)|0\rangle
=\sum_n \,\langle 0|A_\mu(x_1)|n\rangle \: \langle n |
A_\nu(x_2)|0\rangle \; ,
\eeqn
and {\it only} one-(hybrid)-mesons $n$ contribute at 
leading order~\cite{cohen}. Hence there must be a non-empty set of
states $n$ for which $\langle 0|A_\mu(x_1)|n\rangle \: \langle n |
A_\nu(x_2)|0\rangle$ of ${\cal O}(N_c)$. If $A$, $\mu$, $\nu$,
$x_1$ and $x_2$ 
are changed the set of states for which this is true may
change. As these variables are changed, a
specific state $n$ should regularly be part of the set, since there
is nothing special about it.
Hence for a specific particle $\sigma$ with four-momentum 
$p_\sigma$, it must be possible to
choose $A$, $\mu=\nu$ and $x_1=x_2=z$ such that
$\langle 0|A_\mu(x_1)|n\rangle \: \langle n |
A_\nu(x_2)|0\rangle=|\langle \sigma p_\sigma |A_\mu(z)|0\rangle|^2$ 
is ${\cal O}(N_c)$. This implies that 
$\langle \sigma p_\sigma |A_\mu(z)|0\rangle$
is ${\cal O}(\sqrt{N_c})$, as promised.

\subsection{Second preliminary \plabel{appa3}}

The following four-point function is ${\cal O}(N_c^2)$~\cite{coleman}

\beqn
\langle 0|B(x_1)\,C(x_2)\,B(x_3)\,C(x_4)\,
|0\rangle
=\sum_n \,\langle 0|B(x_1)\,C(x_2)|n\rangle \: \langle n |
B(x_3)\,C(x_4)|0\rangle\; ,
\eeqn 
and {\it only} two -(hybrid) meson states $n$ contribute at 
leading order~\cite{coleman}. Similar to the argument for the first
preliminary, it must be possible to choose $B$, $C$, $x_1=
x_3 = (\bx,0)$, $x_2=x_4=0$ and a specific state
$|\sigma_1k_1 \sigma_2k_2\rangle$, such that
$\langle 0|B(x_1)\,C(x_2)|n\rangle \: \langle n |
B(x_3)\,C(x_4)|0\rangle= 
|\langle  0 |\, B(\bx,0) \: C(0)\, |\sigma_1k_1 \sigma_2k_2\rangle|^2$ 
is ${\cal O}(N_c^2)$. Whence
the promised result.

\subsection{Derivation of Eq.~\ref{en} \plabel{appa4}}

All derivations so far left unspecified whether the asymptotic states
were ``in'' or ``out'' states.  Consider the specific case of ``out''
states, and insert a complete set of ``in'' states:

\beqna\plabel{ex}
\lefteqn{\hspace{-14.95cm}\langle \sigma_1k_1 \sigma_2k_2\,\mbox{out}\,|
{A_{\mu}}(z)|0\rangle
=\sum_\sigma\int \frac{d^3 q}{(2\pi)^3}
\:\langle \sigma_1k_1 \sigma_2k_2\,\mbox{out}\, |
\sigma q \,\mbox{in} \rangle\;
\langle \sigma q\,\mbox{in} \, |{A_{\mu}}(z)|0\rangle
\nonumber } \\
+\sum_{\sigma_1'\sigma_2'}\, (1-\frac{\delta_{\sigma_1'\sigma_2'}}{2})\int
\frac{d^3 q_1}{(2\pi)^3} \int \frac{d^3 q_2}{(2\pi)^3}
\:\langle \sigma_1k_1 \sigma_2k_2\,\mbox{out}\,|
\sigma_1' q_1\sigma_2'q_2\,\mbox{in} \rangle\;
\langle \sigma_1' q_1\sigma_2'q_2\,\mbox{in} \,|{A_{\mu}}
(z)|0\rangle \; .
\eeqna
Restrict to $\bk_1+\bk_2  = {\bf 0}$ in this subsection.
The usual large-$N_c$ counting for
$\langle \sigma_1k_1 \sigma_2k_2 | n \rangle$ is order
$1/\sqrt{N_c}$, $1$ (no scattering), 
or $1/\sqrt{N_c}$ for $n$ respectively a
one -, two - or three - (hybrid) meson state~\cite{lebed,cohen}.
The counting for 
$\langle n| {A_{\mu}}(z)|0\rangle$
is respectively $\sqrt{N_c}$, $1$ or $1/\sqrt{N_c}$~\cite{lebed,cohen}.
The product of the countings of the two matrix elements is
$1$, $1$ and $1/N_c$ respectively.
If the asymptotic states contained glueballs the counting of the product
will be lower than $1$.
Hence only one - and two - (hybrid) meson
states contribute in large-$N_c$, as indicated
in Eq.~\ref{ex}. As before, the one-particle states that contribute 
are {\it only} neutral hybrid mesons with the same $J^{PC}$ as
the current $A_{\mu}(0)$.

Connection with the S-matrix is now made by using
$\langle m\, \mbox{out}\, | n \, \mbox{in}\rangle =
\langle m\, \mbox{in}\, | S | n \, \mbox{in}\rangle$~\cite{itzykson}. 
In the second part of Eq.~\ref{ex} write the two-body 
scattering $\langle \sigma_1k_1 \sigma_2k_2\,\mbox{out}\,|
\sigma_1' q_1\sigma_2'q_2\,\mbox{in} \rangle =
\langle \sigma_1k_1 \sigma_2k_2\,\mbox{in}\,|
\sigma_1' q_1\sigma_2'q_2\,\mbox{in} \rangle$ where it was used that 
only no-scattering occurs at ${\cal O}(1)$~\cite{lebed}. 
The latter overlap is 
simply an overlap between free bosonic states in the same basis. It
can be evaluated~\cite{itzykson} and equals

\beqn \plabel{norm}
(2\pi)^6\, (\:
 \delta^3(\bk_1-\bq_1)\, \delta_{\sigma_1\sigma_1'}\,
 \delta^3(\bk_2-\bq_2)\, \delta_{\sigma_2\sigma_2'}
+\delta^3(\bk_1-\bq_2)\, \delta_{\sigma_1\sigma_2'}\,
 \delta^3(\bk_2-\bq_1)\, \delta_{\sigma_2\sigma_1'}\:) \; .
\eeqn
In the first part in Eq.~\ref{ex} introduce the T-matrix,  
defined as the transition from an ``in'' to an ``out'' state

\beqn \plabel{t} 
\langle \sigma_1k_1 \sigma_2k_2\,\mbox{out}\, |
\sigma q \,\mbox{in} \rangle =
i (2\pi)^4 \;\delta^4(k_1 + k_2-q)\;
\langle \sigma_1k_1 \sigma_2k_2\,\mbox{in}\, | T |
\sigma q \,\mbox{in} \rangle \; ,
\eeqn
using that no-scattering does not contribute, given that this is the 
overlap of a two- with a one-particle state; and
employing the definition of the S-matrix in terms of the
``reduced'' T-matrix~\cite{itzykson}. Dropping the ``in'' label
on the R.H.S. of Eq.~\ref{t}, because
$\langle m\, \mbox{in}\, | T | n \, \mbox{in}\rangle =
\langle m\, \mbox{out}\, | T | n \, \mbox{out}\rangle$~\cite{itzykson},
and substituting Eqs.~\ref{norm}-\ref{t} in Eq.~\ref{ex}, yield Eq.~\ref{en}.

\section{Appendix: Why {\it each} term in Eq.~\ref{fk} is 
subleading \plabel{appb}}

This Appendix starts by proving that the angular integration in
Eq.~\ref{fk} can be written as a finite sum. The first two results of
the Paper are then derived. (This is done directly, without first
showing that {\it each} of the terms in Eq.~\ref{fk} is subleading
as discussed in the main text. However, once the first two results are
established, it follows that each of the terms is subleading). The
third result is subsequently derived.

It will be convenient for the derivation to write the integral over
the solid angle $\Omega_{{\bf k}_1}$ in Eq.~\ref{fk}
as a finite sum. This can be done
by performing a partial wave expansion of 
the overlap 
$\langle \sigma_1k_1 \sigma_2k_2| {A_{\mu}}(0)|0\rangle$ by explicitly
considering its Lorentz structure. It is a function of
$k_1$ and $k_2$, or equivalently of $k_1 - k_2$
and $k_1 + k_2$. First consider the Lorentz scalars that can be
built from these two variables: $(k_1-k_2)^2$, $(k_1+k_2)^2$ and
$(k_1-k_2)\cdot(k_1+k_2)$. It is easily shown that the on-shell conditions
$k^2_1 = m_1^2$ and $k^2_2 = m_2^2$ imply that the latter two variables
can be expressed in terms of the first variable. Hence the only
independent Lorentz scalar is $(k_1-k_2)^2$.
Second consider the case where ${A_{\mu}}$ has $J=1$, i.e. is a vector.
The overlap
can be written as a linear combination of
$k_{1\,\mu}$ times a Lorentz scalar, and  $k_{2\,\mu}$ 
times a Lorentz scalar,
since this is the most general structure transforming like a vector. 
Denote the two Lorentz scalars by $\langle \sigma_1 \sigma_2|A(0)\rangle_i\,
((k_1-k_2)^2)$,
with $i=1,2$. Define ${\cal L}^i_\mu\, (k_1,k_2) = k_{i\,\mu}$.
Then the overlap 

\beqn
\langle \sigma_1k_1 \sigma_2k_2| {A_{\mu}}(0)|0\rangle
\equiv \sum_i {\cal L}^i_\mu\, (k_1,k_2) \;
\langle \sigma_1 \sigma_2|A(0)\rangle_i\,((k_1-k_2)^2) \; . \plabel{pwa}
\eeqn
It is evident that the procedure can be performed for arbitrary $J$,
and that appropriate ${\cal L}^i_\mu$ can always be constructed,
with the partial wave $i$ ranging over a finite number of integers. 
Eq.~\ref{pwa} is the promised partial wave expansion of the
overlap for general $J$.
The functions ${\cal L}^i_\mu$ depend purely on kinematical variables
and all dynamical information is contained in the scalar functions
$\langle \sigma_1 \sigma_2|A(0)\rangle_i\,((k_1-k_2)^2)$, which 
depend kinematically 
only
on $(k_1-k_2)^2$. When Eq.~\ref{pwa} is substituted in the two-particle
terms of 
Eq.~\ref{fk} the integral over the solid angle $\Omega_{{\bf k}_1}$
can be written as a finite sum over $i$, as promised, since $(k_1-k_2)^2$
does not depend on the solid angle.

Rewrite Eqs.~\ref{main}-\ref{fk} as

\beqn\plabel{mateq}
W_{BC} = \sum_{\sigma = 1}^{N_\Sigma}\, M_{\,BC\:\sigma}\: V_\sigma +
\sum_{\sigma_1\sigma_2i\; = 1}^{N_\Pi} 
\tilde{M}_{\,BC\;\sigma_1\sigma_2i}\: \tilde{V}_{\sigma_1\sigma_2i} \; , 
\eeqn
where the explicit dependence on the currents $B$ and $C$ are indicated.
The L.H.S. of Eq.~\ref{main} (divided by $N_c$) is

\beqn
W_{BC} \equiv \frac{1}{N_c}
\int^{\infty}_{-\infty} dt\; 
e^{iEt}\; \hat{O}_{\bp}\;    
\int\; d^3x\; d^3y\; e^{\, i\,(\bp\cdot\bx-\bp\cdot\by)}\;
\langle  0 |\: B(\bx,t)\:C(\by,t) \:A_{\mu}(0)\: |0\rangle  \; .
\eeqn
The one-particle matrix elements of Eq.~\ref{fk} (divided by $N_c$) 
are given by
\beqn\plabel{m}
M_{\,BC\:\sigma} \equiv N_c^{\alpha-\frac{1}{2}}\:
2\pi\;\delta(m_\sigma-E)\;\hat{O}_{\bp}\;\langle  0 |\: (\int d^3x\; 
e^{i\bp\cdot\bx}\; {B}(\bx,0){)} \; {C}(0)\:
|\sigma\,{\bf 0}\rangle  \; , 
\eeqn

\vspace{-1cm}
\beqn\plabel{v}
V_\sigma \equiv
 N_c^{-\alpha-\frac{1}{2}}\,\langle \sigma\,{\bf 0}| {A_{\mu}}(0)|0\rangle  
\; , 
\eeqn
where $\alpha$ is a real number specified below. The two-particle
matrix elements of Eq.~\ref{fk} (divided by $N_c$) are

\beqna\plabel{mt}
\lefteqn{\hspace{-13.9cm} \tilde{M}_{\,BC\:\sigma_1\sigma_2i} \equiv 
\frac{1}{N_c}
\frac{1}{(2\pi)^2}\, (1-\frac{\delta_{\sigma_1\sigma_2}}{2})\: K(E) 
\nonumber } \\
\left.\eqntimes\int d\Omega_{{\bf k}_1} 
\;\hat{O}_{\bp}\;\langle  0 |\: (\int d^3x\; 
e^{i\bp\cdot\bx}\; {B}(\bx,0){)} \; {C}(0)\:
|\sigma_1k_1 \sigma_2k_2\rangle 
\;{\cal L}^i_\mu\, (k_1,k_2) \;
\;\right|_{k_1+k_2=({\bf 0},E)} , 
\eeqna

\vspace{-1cm}
\beqn\plabel{vt}
\tilde{V}_{\sigma_1\sigma_2i} \equiv \left.\langle \sigma_1 
\sigma_2|A(0)\rangle_i\,((k_1-k_2)^2)
\;\right|_{k_1+k_2=({\bf 0},E)} ,
\eeqn
using Eq.~\ref{pwa}. As discussed in the main text, each side of
Eq.~\ref{mateq} depends on the kinematical variables $\bp$ and $E$.
For a specific choice of these variables, there are $N_\Sigma \geq 0$
one-particle states $\sigma$ contributing, and
$N_\Pi \geq 0$ two-particle states and partial waves $\sigma_1\sigma_2i$ 
contributing.
It will be useful to think of the group of labels $\sigma_1\sigma_2i$ 
as a single label.
The remainder of the discussion is only of interest if 
it is not the case that $N_\Sigma=N_\Pi=0$.

From the second preliminary
($\langle  0 |\, B(\bx,0) \: C(0)\, |\sigma_1k_1 \sigma_2k_2\rangle$ 
{\it has} to be ${\cal O}(N_c)$) it follows that
$\tilde{M}$ in Eq.~\ref{mt} is $\leq {\cal O}(1)$.
This obtains by noting that
$\hat{O}_{\bp}$ is independent of colour, as is shown at the end of this
Appendix, and that ${\cal L}^i_\mu$ and $K$ are purely kinematical 
functions with no colour dependence.
The possibility that there are accidental
cancellations in the various integrations, which could make $\tilde{M}
<{\cal O}(1)$, is incorporated by indicating that
$\tilde{M}\leq {\cal O}(1)$. The possibility of cancellations will
be taken into account in the derivations below.
By choosing $\alpha$ appropriately, $M$ in Eq.~\ref{m}
is {\it defined} to be exactly ${\cal O}(1)$. 
Hence both $M$ and $\tilde{M}$ are $\leq{\cal O}(1)$.
Evaluate Eq.~\ref{mateq} for $N_\Sigma+N_\Pi$ different currents
$B,\; C$. Since $V$ and $\tilde{V}$ in Eqs.~\ref{v} and~\ref{vt} are 
independent of the choice of currents $B$ and $C$, this amounts to
constructing a matrix equation ${\bf W} = {\sf M}\, {\bf V}$. The 
$N_\Sigma+N_\Pi$ dimensional column vector ${\bf W}$ is built from 
the evaluations of 
$W_{BC}$ for different values of $B,\; C$; the 
$(N_\Sigma+N_\Pi)\times(N_\Sigma+N_\Pi)$ dimensional matrix
${\sf M}$ contains $M_{\,BC\:\sigma}$ and 
$\tilde{M}_{\,BC\:\sigma_1\sigma_2i}$;
and the $N_\Sigma+N_\Pi$ dimensional column vector ${\bf V}$
is built from $V_\sigma$ and $\tilde{V}_{\sigma_1\sigma_2i}$.
Because both $M$ and $\tilde{M}$ are $\leq{\cal O}(1)$ it follows that
each entry of the matrix 
${\sf M}$ is also $\leq{\cal O}(1)$. 
This means that, barring accidental cancellations,
the determinant of ${\sf M}$, det$\,{\sf M}$, which is a sum of
products of the entries of ${\sf M}$, 
is exactly ${\cal O}(1)$, i.e. is non-zero and finite
in the large-$N_c$ limit. Note that even if some of the entries of the
matrix ${\sf M}$ are $<{\cal O}(1)$, it is still possible for
det$\,{\sf M}$ to be exactly ${\cal O}(1)$.
If  det$\,{\sf M} < {\cal O}(1)$ the derivations
below are invalid. This possibility can be excluded by an appropriate
choice of currents.
Since det$\,{\sf M}={\cal O}(1)$ the inverse of ${\sf M}$
exists,
and ${\bf V} = {\sf M}^{-1}\, {\bf W}$. 
Since ${\sf M}^{-1}=$ adj$\, {\sf M}\,/\,$det$\,{\sf M}$,
where adj$\, {\sf M}$ is the adjoint matrix of ${\sf M}$, which is 
a sum of products of entries of ${\sf M}$, it follows that 
the entries of ${\sf M}^{-1} \leq {\cal O}(1)$. Noting from the
main text that the L.H.S. of Eq.~\ref{main} is ${\cal O}(1)$
[strictly speaking, it is $\leq{\cal O}(1)$, since it is only known
that OZI allowed ${\cal O}(N_c)$ contributions are not 
present~\cite{fssr}, and that
the highest order OZI forbidden contribution is 
${\cal O}(1)$~\cite{lebed} if there
are no accidental cancellations], so that
each entry of the vector ${\bf W}\leq{\cal O}(1/N_c)$, it follows from 
${\bf V} = {\sf M}^{-1}\, {\bf W}$ that each entry of the vector
${\bf V}$ is $\leq{\cal O}(1/N_c)$. This implies that
$V$ and $\tilde{V}$ are both $\leq{\cal O}(1/N_c)$. 

It is instructive to study the kinematical variable dependence of
$\tilde{V}$ in Eq.~\ref{vt}, which can only be relevant  
to the discussion if $E$ is above the two-particle threshold 
$m_1+m_2$, since the two-particle term in Eq.~\ref{fk} only has support 
above this threshold. This, together with the constraint  
$k_1+k_2=({\bf 0},E)$ in Eq.~\ref{vt}, and the on-shell
character of the particles, can be shown to imply
that $(k_1-k_2)^2 = 2\;(m_1^2+m_2^2)-E^2 \in (-\infty,\,(m_1-m_2)^2\,]$.
Using $\tilde{V}\leq{\cal O}(1/N_c)$, and Eqs.~\ref{pwa} and~\ref{vt},
it follows that $\langle \sigma_1k_1 \sigma_2k_2| {A_{\mu}}(0)|0\rangle
\leq{\cal O}(1/N_c)$ with the constraint 
that $(k_1-k_2)^2 \in (-\infty,(m_1-m_2)^2\,]$.
However, it is possible to show by only considering the on-shell
nature of the particles, that the same constraint holds. Hence the
constraint adds no new information, and is dropped henceforth. Thus

\beqn\plabel{ar1}
\langle \sigma_1k_1 \sigma_2k_2| {A_{\mu}}(z)|0\rangle
= e^{i (k_1+k_2)\cdot z} 
\langle \sigma_1k_1 \sigma_2k_2| {A_{\mu}}(0)|0\rangle
\leq{\cal O}(1/N_c) \; ,
\eeqn
using space-time translational invariance 
$A_{\mu}(z) = e^{iP\cdot z} A_{\mu}(0)\, e^{-iP\cdot z}$,
with $P^\nu$ the QCD four-momentum operator. 
It is evident that equality in Eq.~\ref{ar1} would have been attained
were it not for the possibility of accidental cancellations. These
cancellations can be eliminated by an appropriate choice of the 
currents. Whence the result in Eq.~\ref{result1}.

The observation that $V\leq{\cal O}(1/N_c)$,
together with the first preliminary
($\langle \sigma\,{\bf 0}| {A_{\mu}}(0)|0\rangle$ {\it has} to be
${\cal O}(\sqrt{N_c})$), implies from Eq.~\ref{v} that
$\alpha\geq 1$. Since $M$ was defined to be ${\cal O}(1)$ this implies
from Eq.~\ref{m} that 
$\langle  0 |\: (\int d^3x\; 
e^{i\bp\cdot\bx}\; {B}(\bx,0){)} \; {C}(0)\:
|\sigma\,{\bf 0}\rangle$ is $\leq {\cal O}(1/\sqrt{N_c})$, noting that
$m_\sigma$ is ${\cal O}(1)$~\cite{cohen}.
Inverting the Fourier transform, it follows that
$\langle  0 |\: B(\bx,0) \; C(0)\:
|\sigma\,{\bf 0}\rangle \leq {\cal O}(1/\sqrt{N_c})$. Using
space-time translational invariance analogous to Eq.~\ref{ar1}

\beqn
\langle  0 |\: B(\bx,t) \; C(\by,t)\:|\sigma\,{\bf 0}\rangle =
e^{-im_\sigma t}\;
\langle  0 |\: B(\bx-\by,0) \; C(0)\:|\sigma\,{\bf 0}\rangle
\leq {\cal O}(1/\sqrt{N_c}) \; .
\eeqn
From the same arguments as those below Eq.~\ref{ar1},
the result in Eq.~\ref{result2} is deduced.

It remains to prove that Eq.~\ref{result3} can be deduced from 
Eq.~\ref{en}. The proof is analogous to the proof already given in 
this Appendix, and the various steps are outlined. The notation
of the vectors and matrices will be the same except that the label
$BC$ will be replaced by the label $A$. Call the L.H.S. of Eq.~\ref{en}
$W_A$. The first and second terms in long brackets on the R.H.S.
of Eq.~\ref{en} are called $M_{A\sigma}$ and $V_\sigma$ respectively.
Eq.~\ref{en} is then of the form of Eq.~\ref{mateq} with $N_\Pi=0$.
From the first preliminary
($\langle \sigma \, {\bf 0}|{A_{\mu}}(z)|0\rangle$
{\it has} to be
${\cal O}(\sqrt{N_c})$), it follows that $M_{A\sigma}$ is 
${\cal O}(1)$. Evaluate Eq.~\ref{mateq} for $N_\Sigma$ different
currents $A$. This again gives a matrix equation 
 ${\bf W} = {\sf M}\, {\bf V}$. Barring accidental cancellations,
det$\, {\sf M} = {\cal O}(1)$, so that its inverse exists.
 From Eq.~\ref{result1} 
${\bf W} \leq {\cal O}(1/N_c)$, and together with ${\sf M}^{-1}
\leq {\cal O}(1)$, the equation 
${\bf V} = {\sf M}^{-1}\, {\bf W}$ implies that 
${\bf V} \leq {\cal O}(1/N_c)$. Along the same lines as before this
establishes the result in Eq.~\ref{result3}.

It is lastly outlined why $\hat{O}_{\bp}$ does not depend on colour.
This is done by following the derivation of $\hat{O}_{\bp}$ in the
Appendix of Ref.~\cite{fssr}, employing the notations of that reference.
Colour appears when the 
(anti)commutators are evaluated, e.g. as $\delta^{ab}$ 
in $\{\dot{\psi}_\xi^a(\bx,t),\bar{\psi}_\zeta^b(\by,t)\}=
-\delta^{ab}\vec{\gamma}_{\xi\zeta}\cdot\vec{\partial}_{\bx}
\delta^3(\bx-\by)$.
The colour and Dirac indices in the commutators are then
contracted with the remaining quark and gluon fields and subsumed
in $f_\mu(x,y,z)$. The construction of $\hat{O}_{\bp}$ only depends
on the number of derivatives acting on $\delta^3(\bx-\by)$ and not
on $f_\mu(x,y,z)$, making it independent of colour, as promised.
Let's give an example of how this observation is used above. 
Suppose $\hat{O}_{\bp} =
\partial / \partial \bp$, then in Eqs.~\ref{m} and~\ref{mt} it occurs
as $\hat{O}_{\bp} \int d^3 x \exp(i\bp\cdot\bx)\, g(\bx) = 
i\bx \int d^3 x \exp(i\bp\cdot\bx)\, g(\bx)$. It is evident that
$\hat{O}_{\bp}$ does not affect the large-$N_c$ counting of the 
function $g$.

\end{document}